\documentstyle[twoside]{article}

\catcode`\@=11
\long\def\@makefntext#1{
\protect\noindent \hbox to 3.2pt {\hskip-.9pt  
$^{{\eightrm\@thefnmark}}$\hfil}#1\hfill}		

\def\@makefnmark{\hbox to 0pt{$^{\@thefnmark}$\hss}}	
	
\def\ps@myheadings{\let\@mkboth\@gobbletwo
\def\@oddhead{\hbox{}
\rightmark\hfil\eightrm\thepage}   
\def\@oddfoot{}\def\@evenhead{\eightrm\thepage\hfil
\leftmark\hbox{}}\def\@evenfoot{}
\def\sectionmark##1{}\def\subsectionmark##1{}}

\oddsidemargin=\evensidemargin
\newcounter{sectionc}\newcounter{subsectionc}\newcounter{subsubsectionc}
\renewcommand{\section}[1] {\vspace{12pt}\addtocounter{sectionc}{1} 
\setcounter{subsectionc}{0}\setcounter{subsubsectionc}{0}\noindent 
	{\tenbf\thesectionc. #1}\par\vspace{5pt}}
\renewcommand{\subsection}[1] {\vspace{12pt}\addtocounter{subsectionc}{1} 
	\setcounter{subsubsectionc}{0}\noindent 
	{\bf\thesectionc.\thesubsectionc. {\kern1pt \bfit #1}}\par\vspace{5pt}}
\renewcommand{\subsubsection}[1] {\vspace{12pt}\addtocounter{subsubsectionc}{1}
	\noindent{\tenrm\thesectionc.\thesubsectionc.\thesubsubsectionc.
	{\kern1pt \tenit #1}}\par\vspace{5pt}}
\newcommand{\nonumsection}[1] {\vspace{12pt}\noindent{\tenbf #1}
	\par\vspace{5pt}}

\newcounter{appendixc}
\newcounter{subappendixc}[appendixc]
\newcounter{subsubappendixc}[subappendixc]
\renewcommand{\thesubappendixc}{\Alph{appendixc}.\arabic{subappendixc}}
\renewcommand{\thesubsubappendixc}
	{\Alph{appendixc}.\arabic{subappendixc}.\arabic{subsubappendixc}}

\renewcommand{\appendix}[1] {\vspace{12pt}
        \refstepcounter{appendixc}
        \setcounter{figure}{0}
        \setcounter{table}{0}
        \setcounter{lemma}{0}
        \setcounter{theorem}{0}
        \setcounter{corollary}{0}
        \setcounter{definition}{0}
        \setcounter{equation}{0}
        \renewcommand{\thefigure}{\Alph{appendixc}.\arabic{figure}}
        \renewcommand{\thetable}{\Alph{appendixc}.\arabic{table}}
        \renewcommand{\theappendixc}{\Alph{appendixc}}
        \renewcommand{\thelemma}{\Alph{appendixc}.\arabic{lemma}}
        \renewcommand{\thetheorem}{\Alph{appendixc}.\arabic{theorem}}
        \renewcommand{\thedefinition}{\Alph{appendixc}.\arabic{definition}}
        \renewcommand{\thecorollary}{\Alph{appendixc}.\arabic{corollary}}
        \renewcommand{\theequation}{\Alph{appendixc}.\arabic{equation}}
        \noindent{\tenbf Appendix \theappendixc #1}\par\vspace{5pt}}
\newcommand{\subappendix}[1] {\vspace{12pt}
        \refstepcounter{subappendixc}
        \noindent{\bf Appendix \thesubappendixc. {\kern1pt \bfit #1}}
	\par\vspace{5pt}}
\newcommand{\subsubappendix}[1] {\vspace{12pt}
        \refstepcounter{subsubappendixc}
        \noindent{\rm Appendix \thesubsubappendixc. {\kern1pt \tenit #1}}
	\par\vspace{5pt}}

\newcommand{\textlineskip}{\baselineskip=13pt}
\newcommand{\smalllineskip}{\baselineskip=10pt}

\def\eightcirc{
\begin{picture}(0,0)
\put(4.4,1.8){\circle{6.5}}
\end{picture}}
\def\eightcopyright{\eightcirc\kern2.7pt\hbox{\eightrm c}} 

\newcommand{\copyrightheading}[1]
	{\vspace*{-2.5cm}\smalllineskip{\flushleft
	{\footnotesize Modern Physics Letters A, #1}\\
	{\footnotesize $\eightcopyright$\, World Scientific Publishing
	 Company}\\
	 }}

\def\abstracts#1#2#3{{
	\centering{\begin{minipage}{4.5in}\baselineskip=10pt\footnotesize
	\parindent=0pt #1\par 
	\parindent=15pt #2\par
	\parindent=15pt #3
	\end{minipage}}\par}} 

\newcommand{\bibit}{\nineit}

\renewenvironment{thebibliography}[1]
	{\frenchspacing
	 \ninerm\baselineskip=11pt
	 \begin{list}{\arabic{enumi}.}
        {\usecounter{enumi}\setlength{\parsep}{0pt}     
	 \setlength{\leftmargin 12.7pt}{\rightmargin 0pt} 
         \setlength{\itemsep}{0pt} \settowidth
	{\labelwidth}{#1.}\sloppy}}{\end{list}}

\newcounter{itemlistc}
\newcounter{romanlistc}
\newcounter{alphlistc}
\newcounter{arabiclistc}

\newcommand{\fcaption}[1]{
        \refstepcounter{figure}
        \setbox\@tempboxa = \hbox{\footnotesize Fig.~\thefigure. #1}
        \ifdim \wd\@tempboxa > 5in
           {\begin{center}
        \parbox{5in}{\footnotesize\smalllineskip Fig.~\thefigure. #1}
            \end{center}}
        \else
             {\begin{center}
             {\footnotesize Fig.~\thefigure. #1}
              \end{center}}
        \fi}

\newcommand{\tcaption}[1]{
        \refstepcounter{table}
        \setbox\@tempboxa = \hbox{\footnotesize Table~\thetable. #1}
        \ifdim \wd\@tempboxa > 5in
           {\begin{center}
        \parbox{5in}{\footnotesize\smalllineskip Table~\thetable. #1}
            \end{center}}
        \else
             {\begin{center}
             {\footnotesize Table~\thetable. #1}
              \end{center}}
        \fi}

\def\@citex[#1]#2{\if@filesw\immediate\write\@auxout
	{\string\citation{#2}}\fi
\def\@citea{}\@cite{\@for\@citeb:=#2\do
	{\@citea\def\@citea{,}\@ifundefined
	{b@\@citeb}{{\bf ?}\@warning
	{Citation `\@citeb' on page \thepage \space undefined}}
	{\csname b@\@citeb\endcsname}}}{#1}}

\newif\if@cghi
\def\cite{\@cghitrue\@ifnextchar [{\@tempswatrue
	\@citex}{\@tempswafalse\@citex[]}}
\def\citelow{\@cghifalse\@ifnextchar [{\@tempswatrue
	\@citex}{\@tempswafalse\@citex[]}}
\def\@cite#1#2{{$\null^{#1}$\if@tempswa\typeout
	{IJCGA warning: optional citation argument 
	ignored: `#2'} \fi}}

\def\@refcitex[#1]#2{\if@filesw\immediate\write\@auxout
	{\string\citation{#2}}\fi
\def\@citea{}\@refcite{\@for\@citeb:=#2\do
	{\@citea\def\@citea{, }\@ifundefined
	{b@\@citeb}{{\bf ?}\@warning
	{Citation `\@citeb' on page \thepage \space undefined}}
	\hbox{\csname b@\@citeb\endcsname}}}{#1}}

\def\@refcite#1#2{{#1\if@tempswa\typeout
        {IJCGA warning: optional citation argument
	ignored: `#2'} \fi}}

\def\refcite{\@ifnextchar[{\@tempswatrue
	\@refcitex}{\@tempswafalse\@refcitex[]}}

\def\pmb#1{\setbox0=\hbox{#1}
	\kern-.025em\copy0\kern-\wd0
	\kern.05em\copy0\kern-\wd0
	\kern-.025em\raise.0433em\box0}

\def\fnm#1{$^{\mbox{\scriptsize #1}}$}
\def\fnt#1#2{\footnotetext{\kern-.3em
	{$^{\mbox{\scriptsize #1}}$}{#2}}}

\def\fpage#1{\begingroup
\voffset=.3in
\thispagestyle{empty}\begin{table}[b]\centerline{\footnotesize #1}
	\end{table}\endgroup}

\def\runninghead#1#2{\pagestyle{myheadings}
\markboth{{\protect\footnotesize\it{\quad #1}}\hfill}
{\hfill{\protect\footnotesize\it{#2\quad}}}}
\font\tenrm=cmr10
\font\tenit=cmti10 
\font\tenbf=cmbx10
\font\bfit=cmbxti10 at 10pt
\font\ninerm=cmr9
\font\nineit=cmti9

\font\eightrm=cmr8

\def\qed{\hbox{${\vcenter{\vbox{			
   \hrule height 0.4pt\hbox{\vrule width 0.4pt height 6pt
   \kern5pt\vrule width 0.4pt}\hrule height 0.4pt}}}$}}

\begin{document}

\runninghead{Lorentz Multiplet Structure of Baryons
$\ldots$} {Lorentz Multiplet Structure of Baryons $\ldots$}

\normalsize\textlineskip
\thispagestyle{empty}
\setcounter{page}{1}

\copyrightheading{}			

\vspace*{0.88truein}

\fpage{1}
\centerline{\bf LORENTZ MULTIPLET STRUCTURE OF BARYON
SPECTRA} 
\vspace*{0.035truein}
\centerline{\bf AND RELATIVISTIC DESCRIPTION}
\vspace*{0.37truein}
\centerline{\footnotesize M. KIRCHBACH
\footnote{E-mail: mariana@kph.uni-mainz.de}}
\vspace*{0.015truein}
\centerline{\footnotesize\it Institut f\"ur Kernphysik,} 
\baselineskip=10pt
\centerline{\footnotesize\it J. Gutenberg 
Universit\"at Mainz, D-55099 Mainz, Germany}


\vspace*{0.21truein}
\abstracts{
The pole positions of the various baryon resonances are known
to reveal well--pronounced clustering, so--called H\"ohler clusters.
{}For nonstrange baryons the H\"ohler clusters are shown to be 
identical to Lorentz multiplets of the type $\lbrace j,j\rbrace \otimes 
\lbrack\lbrace 1/2,0\rbrace \oplus \lbrace 0,1/2\rbrace \rbrack $
with $j $ being a half--integer. For the $\Lambda $ hyperons below 
1800 MeV these clusters are shown to be of the type 
$\lbrace 1,0\rbrace \oplus \lbrace 0,1\rbrace\otimes 
\lbrack\lbrace 1/2,0\rbrace \oplus \lbrace 0,1/2\rbrace \rbrack $
while above 1800 MeV they are parity duplicated $\lbrace J,0\rbrace \oplus
\lbrace 0, J \rbrace $ higher--spin (Weinberg--Ahluwalia) states. 
Therefore, for $\Lambda $ hyperons the 
restoration of chiral symmetry takes place above 1800 MeV.
{}Finally, it is demonstrated that the description of spin--3/2 
particles in terms of a 2nd rank antisymmetric Lorentz tensor with Dirac 
spinor components does not contain any off--shell parameters and 
avoids the main difficulties of the Rarita--Schwinger description based 
upon a 4--vector with Dirac spinor components.}{}{}



\vspace*{1pt}\textlineskip	
\section{Introduction }	
\vspace*{-0.5pt}
\noindent
A recent analysis of the pole positions of various baryon 
resonances  ($N^*$) with masses below $\sim $ 2500 MeV performed 
by H\"ohler et al.\cite{Hoehler} reveals a well--pronounced clustering.
This is quite a surprising result in that so far it was not 
anticipated by any model or theory.\cite{Nefkens}
Subsequently, these baryon clusters will be referred to as
`H\"ohler clusters' according to a suggestion of Nefkens.\cite{Nefkens}
In conjunction with this observation, the symmetry of all reported nonstrange 
baryon excitations with masses below 2500 MeV was re--analyzed\cite{Kirchbach} 
and shown to be governed by SL(2,C) $\otimes $SU(2)$_I$. As long as the 
group SL(2,C) is the universal covering of the Lorentz group, the new 
classification scheme for baryons is  determined by O(1,3)$\otimes $SU(2)$_I$ 
rather than by O(3)$\otimes $ SU(6) which constantly has been used since the 
invention of the naive three flavor quark model till now.

The O(1,3)$\otimes $SU(2)$_I$ symmetry indicates that
the spin--orbital correlation between quarks is much stronger than the 
spin--flavor one. For that reason, the production 
of relativistic multiplets of the type 
$\lbrace j,j\rbrace \otimes 
\lbrack \lbrace 1/2,0\rbrace \oplus \lbrace 0, 1/2\rbrace\rbrack $
(with $2j+1$ even), and thereby of resonance clusters, 
seems to be favored in nature over that of isolated 
$\lbrace J,0\rbrace \oplus \lbrace 0, J\rbrace $ 
higher--spin states. 
It is one of the purposes of the present 
investigation to point out an equivalence between the  Lorentz 
multiplets and the H\"ohler clusters.

The existence of O(1,3)$\otimes $SU(2)$_I$ symmetry for baryons 
encountered in Ref.~\refcite{Kirchbach} might be quite advantageous 
in solving the problem of the relativistic description of their 
spectra.\fnm{a}\fnt{a}{The main problem with the relativization 
of the naive three--flavor quark model is that the Lorentz boost is 
responsible for transitions between the various O(3) multiplets and
mixes different representations of the group O(3)$\otimes $SU(6).}
Indeed, as long as the projection operator onto each given representation 
of the Lorentz group is a well defined mathematical object, one 
can immediately write down the relativistic equation of motion
describing the resonance cluster as a whole
and obtain the corresponding propagator.
In evaluating Feynman graphs, describing physical processes 
such as meson production off protons, it appears quite reasonable,
therefore, to use a propagating  H\"ohler cluster
as an intermediate  state, rather than an isolated higher--spin state.

Lorentz multiplets of the type observed in the baryon spectra 
were first considered  systematically by Weinberg in connection with 
relativistic equations of motion for particles with any 
spin.\cite{Weinberg} There, the representations 
$\lbrace J, 0\rbrace \oplus \lbrace 0, J\rbrace $
of the Lorentz group have been mapped onto multiplets of 
the type $\lbrace {1\over 2}l,{1\over 2} l \rbrace \otimes 
\lbrack\lbrace 1/2,0\rbrace \oplus \lbrace 0, 1/2\rbrace \rbrack $
with integer $l$. Such multiplets neccessarily contain $(2l+1)$ states 
with spins $J$ ranging from $J=1/2$ to $J=l+1/2 $. 
Both representations mentioned above are described in terms of 
totally symmetric ${2l+2}$--rank  multispinors. 
The isolated field of maximal spin is then obtained after eliminating 
all the redundant fields with lower spins by means of a suitably chosen 
set of subsidiary conditions. It is well known 
that all equations of motion for particles with arbitrary 
spins based upon $\lbrace {1\over 2}l,{1\over 2}l \rbrace \otimes 
\lbrack\lbrace 1/2,0\rbrace \oplus \lbrace 0, 1/2\rbrace \rbrack $
representations possess a pathological property. 
When placed in an electromagnetic filed, the higher--spin field ($N^*$) 
can propagate with superluminal velocity and violate 
causality.\cite{superluminal,Goldman} In addition, arbitrary parameters 
arise both in the propagator and the $NN^*\gamma $ vertex when the particle 
is treated off--mass shell.\cite{Bennm}  
Recent progress in the description of the 
$\lbrace J,0\rbrace \oplus\lbrace 0,J\rbrace $ states 
avoiding the first difficulty was achieved by Ahluwalia et al.,\cite{dva} 
where the states considered were treated as multicomponent bi--vectors 
rather than as $(2J+1)$--rank multispinors. The covariant behavior of these 
states with respect to Lorentz transformations was then ensured by 
construction in terms of the explicit representation of the 
boost operation within the corresponding $2(2J+1)$ dimensional spaces.
In the following, isolated $\lbrace J,0\rbrace \oplus\lbrace 0,J\rbrace $ 
states will sometimes be referred to as Weinberg--Ahluwalia states to
distinguish them from H\"ohler clusters. 

The most popular equation of motion for a particle with higher
spin is the one built upon the 
$\lbrace 1/2,1/2\rbrace \otimes \lbrack\lbrace 1/2,0\rbrace \oplus 
\lbrace 0,1/2\rbrace \rbrack $ representation,
known as the Rarita--Schwinger (RS) equation.\cite{Lurie}
It describes a spin--3/2 particle together with two  further 
spin--1/2 fields (that need be eliminated by a set of suitably constructed 
subsidiary conditions) in terms a Lorentz vector with Dirac spinor 
components. Here we compare the RS equation to the one based on the
$\lbrack\lbrace 1,0\rbrace \oplus \lbrace 0, 1\rbrace \rbrack
\otimes
\lbrack\lbrace 1/2,0\rbrace \oplus \lbrace 0, 1/2\rbrace \rbrack $
representation space that describes a two--resonance cluster 
with spins $3/2$, and $1/2$, respectively, in terms of a totally
antisymmetric Lorentz tensor with Dirac spinor components. 
We point out that the exclusion of the spin--1/2 component
by means of a subsidiary condition constructed by Fushchich and 
Nikitin\cite{FuNi} directly from the squared Pauli--Lubanski vector 
leads to an equation of motion for the spin--3/2 resonance that avoids
the main difficulties of the Rarita--Schwinger ansatz.

The paper is organized as follows. In the next two sections
we briefly review for completeness the structure of nonstrange 
baryon spectra in terms of O(1,3)$\otimes $SU(2)$_I$ representations
and present the classification of the $\Lambda $ hyperon excitations
with special emphasis on the parity doublet patterns.  
In section IV we outline the construction of `resonance cluster'
propagators and present the new formalism for spin-3/2 fields. 
The paper closes with a short summary.

\vspace*{1pt}\textlineskip	
\section{Clustering of Baryon Spectra}	
\vspace*{-0.5pt}
\noindent
As long as the $\lbrace j,j\rbrace $ multiplets emerge in the O(4) 
symmetric Coulomb problem,\cite{EllDo} where they correspond to a 
principal quantum number $n=2j+1$, the excitations of 
each single baryon isomultiplet with a $\pi N$ decay mode
described in Ref.~\refcite{Kirchbach} appear amazingly patterned 
after the spectrum of the hydrogen atom with even principal 
quantum numbers $n=2,4,6$. In the O(3) reduction, the  
$\lbrace j,j\rbrace $ multiplets appear filled with states of (integer) 
angular momentum $l$ taking the values $l=0, ..., n-1$. 
All angular momenta contained in such representations have either natural
or unnatural parity. Coupling a Dirac spinor to $l$ is standard and leads 
to states with total spin $\vec{J}=\vec{l}\otimes \vec {1\over 2}$. 
For n=2 one finds, therefore, $J = {1\over 2}, {1\over 2}$, 
and ${3\over 2}$.\fnm{b}\fnt{b}{The parity $(-1)^{L+1}$ of a single 
$\pi N$ resonance $L_{2I,2J}$ in standard notation\cite{PDG} is determined 
in the present classification by either $(-1)^l$ or
$(-1)^{l+1}$, depending on whether the parity of the intrinsic
orbital angular momentum is natural
or unnatural. In the present notation, L takes the values of either 
$L=|l-1|, (l+1)$ for natural, or $L=l$ for unnatural parities.}
The parity of the latter states is exclusively 
determined by that of the underlying O(3) states. 
In case of natural parities, i.e., $0^+, 1^-$,  one finds the sequence
$J = {1\over 2}^+, {1\over 2}^-$, and $ {3\over 2}^-$, while for unnatural 
parities, i.e., $0^-, 1^+$, one has $J={1\over 2}^- , 
{1\over 2}^+$, and ${3\over 2}^+$, correspondingly. From this simple 
example it is seen that the parity of the state with the maximal spin 
allows one to reconstruct the parity (natural or unnatural) 
of the members of the Coulomb multiplet. Now, in considering the 
lowest excitations of the nucleon, the $\Lambda $, and the $\Delta $ 
resonances, one always finds the states P$_{2I, 1}$, S$_{2I, 1}$, and 
D$_{2I, 3}$. This means that the parity of the corresponding intrinsic 
orbital angular momenta $l$ is natural and that these baryons are built 
upon a $0^+$ vacuum as spontaneously selected in the Nambu-Goldstone mode 
of chiral symmetry.\cite{Coleman} 

Now, in considering the isospin--3/2 Lorentz multiplet with
n=4, one finds all the required seven $\Delta $ resonances 
with spins ranging from ${1\over 2}^-$ to ${7\over 2}^+$ 
to be concentrated in the tiny mass region between 1900 MeV and 
1950 MeV. The positive parity of the highest spin $F_{37} $ state 
means that it has an intrinsic orbital angular momentum
$l=3^+$ of unnatural parity.
Therefore, the second $\Delta $ resonance cluster
is built upon a $0^-$ vacuum and here chiral symmetry is restored. 
Within the O(1,3)$\otimes $SU(2)$_I$ scheme, the $\Delta $ spectrum 
below 2 GeV appears complete.
The F$_{37}$ state has to be paralleled in the nucleon sector 
by a still-missing F$_{17}$ resonance with a mass around 1700 MeV. 
The second cluster of nucleon resonances is consequently  also expected to be 
built upon a vacuum of unnatural parity, and it is there where chiral symmetry 
restoration for nucleons will take place.

In comparing the n=6 nucleon and $\Delta $ states, four more missing 
resonances are predicted. These are the H$_{1, 11}$, P$_{31}$, P$_{33}$, 
and D$_{33}$ states with masses between 2200 and 2400 MeV.
In summary, five new, still unobserved non--strange resonances have been 
predicted in Ref.~\refcite{Kirchbach}

Now, the comparison between the O(1,3)$\otimes $SU(2)$_I$ multiplets
and the H\"ohler clusters performed in Table 1 reveals their identity.
{}For example, the $\Delta $ cluster at 1900 MeV is in fact nothing else 
but the isospin--3/2 $\lbrace 3/2,3/2\rbrace \otimes \lbrack 
\lbrace 1/2,0\rbrace \oplus 
\lbrace 0,1/2 \rbrace\rbrack $ Lorentz multiplet.  
H\"ohler clusters are, therefore, Lorentz multiplets.

\begin{table}[htbp]
\tcaption{Correspondence between H\"ohler clusters and Lorentz multiplets.
 The missing resonances predicted here have been labeled by `$ ms $'. For 
 other notations, see text.}
\medskip
\centerline{\footnotesize\smallskip
\begin{tabular}{lll}
\hline
\\
$\bf\mbox{L} _{2I,2J} $ {\bf States} & {\bf Pole (MeV)} & 
{\bf Lorentz Multiplet} \\
\\ 
\hline
~\\
S$ _{11} $, P$ _{11} $, P$ _{13} $, D$ _{13} $, & (1665 $\pm $25) & 
$ \lbrace\frac{3}{2}, \frac{3}{2}\rbrace \otimes $
$ \lbrack \lbrace \frac{1}{2}, 0 \rbrace \oplus $
$ \lbrace 0,\frac{1}{2}\rbrace \rbrack $ \\
D$ _{15} $, F$ _{15} $, F$ _{17}^{ms} $ 
& -(55 $\pm $ 15)i & $ \otimes \lbrace \frac{1}{2}\rbrace _I $ \\
 & & ( n=4)\\
~\\
\hline
~\\
S$ _{11} $, P$ _{11} $, P$_{13}$, D$ _{13} $,  &
(2110 $\pm $ 50) & $ \lbrace \frac{5}{2},\frac{5}{2}\rbrace \otimes $
$ \lbrack \lbrace \frac{1}{2}, 0 \rbrace \oplus \lbrace 0, \frac{1}{2} $
$ \rbrace \rbrack $\\
D$ _{15} $, F$_{15}$, F$ _{17} $, G$ _{17} $, &
-(180 $\pm $50)i & $\otimes \lbrace \frac{1}{2} \rbrace _I $   \\
 G$ _{19} $, H$ _{19} $, H$ _{1,11}^{ms} $ &  & (n=6) \\
~\\
\hline
~\\
S$ _{31} $, P$ _{31} $, P$ _{33} $, D$ _{33} $, & (1820 $\pm $30)&
$ \lbrace \frac{3}{2}, \frac{3}{2}\rbrace \otimes $ 
$ \lbrack \lbrace \frac{1}{2},0 \rbrace \oplus \lbrace 0,\frac{1}{2} $
$\rbrace\rbrack  $\\
D$ _{35} $, F$ _{35} $, F$ _{37} $ & -(120 $\pm $30)i &
$\otimes \lbrace \frac{3}{2} \rbrace_I $  \\
 & & (n=4)\\
~\\
\hline
~\\
S$ _{31} $, P$ _{31}^{ms} $, P$ _{33}^{ms} $, D$_{33}^{ms} $, &  &
$\lbrace \frac{5}{2}, \frac{5}{2} \rbrace \otimes \lbrack $
$ \lbrace \frac{1}{2},0 \rbrace \oplus \lbrace 0,\frac{1}{2} $
$\rbrace\rbrack  $ \\
D$ _{35} $, F$ _{35}$, F$ _{37}$, G$ _{37} $ &
less established & $\otimes \lbrace \frac{3}{2}\rbrace _I $ \\
G$ _{39} $, H$_{39}$, H$ _{3,11} $ & & (n=6) \\
~\\
\hline
~\\
S$ _{01} $(1800) , P$ _{01} $(1810) &  & 
$ \lbrace 0^\pm\rbrace \otimes \lbrack \lbrace \frac{1}{2},0 \rbrace $ 
$ \oplus \lbrace 0,\frac{1}{2}\rbrace \rbrack $\\
D$ _{05} $(1830), F$ _{05} $(1820) &  &
$\lbrace 0^\pm \rbrace \otimes\lbrack \lbrace \frac{5}{2},0 \rbrace $ 
$\oplus \lbrace 0, \frac{5}{2}\rbrace \rbrack $\\
P$ _{03} $ (1890), D$ _{03}^{ms} $(2000?) &  &
$\lbrace 0^\pm\rbrace \otimes\lbrack \lbrace \frac{3}{2},0 \rbrace $ 
$\oplus \lbrace 0,\frac{3}{2}\rbrace \rbrack $\\
G$ _{07} $(2100), F$ _{07} $(2020) &  &
$\lbrace 0^\pm\rbrace \otimes\lbrack \lbrace \frac{7}{2},0 \rbrace $ 
$\oplus \lbrace 0,\frac{7}{2}\rbrace \rbrack $\\
~\\
\hline
\end{tabular}}
\end{table}

\vspace*{1pt}\textlineskip	
\section{Parity Doublets in the $\Lambda $ Hyperon Spectrum}	
\vspace*{-0.5pt}
\noindent
In Ref.~\refcite{Kirchbach} it was shown that none of the nonstrange 
resonances is exact parity duplicated. Indeed, exact fermion parity 
doublets must originate from underlying equal intrinsic orbital 
angular momenta of opposite parities. All the observed nearly mass 
degenerate nonstrange baryons of equal spins and opposite parities have, 
however, intrinsic orbital angular momenta differing by one unit.
{}For example, the states N$^*$(1675) and N$^*$(1680) with 
$J^P={5\over 2}^-$, and $J^P={5\over 2}^+$, respectively, 
do not really form a pair because they belong to one and the same
Lorentz multiplet (that corresponding to n=4), and their
intrinsic orbital angular momenta and parities are $l=2^-$ and $l=3^+$, 
respectively. On the contrary, in the present section it is shown that
excited hyperons at comparatively high energies
clusterize into exact parity doublets.

In Ref.~\refcite{Kirchbach} only the lowest $\Lambda $ hyperon excitations
were considered. There, the first excited S$_{01}$, P$_{01}$, and 
D$_{03}$ states were organized into the isosinglet
$\lbrace 1/2,1/2\rbrace \otimes  
\lbrack\lbrace 1/2,0\rbrace \oplus
\lbrace 0,1/2\rbrace \rbrack $  representation.
The splitting between these S$_{01}$ and D$_{03}$ resonances
reaches  175 MeV, and is comparable to
the gap of about 150 MeV between their mass average 
and the P$_{01}$ state. This observation indicates that a further 
reduction of the four dimensional Lorentz multiplet into the trivial and 
the 3-vector representations is useful:
\begin{equation}
\lbrace {1\over 2},{1\over 2} \rbrace  
\longrightarrow \lbrace 0\rbrace \oplus 
\lbrack \lbrace 1,0\rbrace \oplus \lbrace 0,1\rbrace  
\rbrack \, .
\label{Maxwell}
\end{equation}
The first excited P$_{01}$ state is now attached to 
\begin{equation}
\Lambda (1600;1/2^+) \simeq \lbrace 0^+ \rbrace \otimes                          
\lbrack\lbrace {1\over 2},0\rbrace \oplus
\lbrace 0, {1\over 2} \rbrace \rbrack \, ,
\label{L1405}
\end{equation}
while the first S$_{01}$ and D$_{03}$ states are combined into
\begin{equation} 
\Lambda (1405;{1\over 2}^-) \quad \mbox{and}\quad 
\Lambda (1520;{3\over 2}^-)
\simeq \lbrack \lbrace 1,0\rbrace \oplus \lbrace 0,1\rbrace  
\rbrack
\otimes                          
\lbrack\lbrace {1\over 2},0\rbrace \oplus
\lbrace 0, {1\over 2} \rbrace \rbrack \, .
\label{L1520}
\end{equation}
The $\Lambda (1600) $ resonance can therefore be considered as an isolated 
second spin--1/2$^+$ $\Lambda $ hyperon that parallels the Roper resonance 
of the nucleon. From the last equation it is further seen that the first 
isosinglet spin-$1/2^-$ resonance has an intrinsic angular momentum  $1^-$ 
and cannot be considered as the parity partner to $\Lambda (1116)$ of 
$0^+$ intrinsic angular momentum. The $\lbrace 1,0\rbrace \oplus \lbrace 0,1
\rbrace $ pattern appears once more in the next resonance cluster including 
the $\Lambda (1670;1/2^-)$ and $\Lambda (1690;3/2^-)$ states. Here again, the 
second S$_{01}$ excitation is still not a parity partner to $\Lambda (1116)$.
At 1800 MeV, however, one immediately encounters well-pronounced parity
doublet clusters according to
\begin{eqnarray} 
\Lambda (1800;1/2^-) &\simeq&  \lbrace 0^-\rbrace 
\otimes \lbrack \lbrace {1\over 2},0\rbrace \oplus \lbrace 0,{1\over 2}
\rbrace\rbrack  ,\nonumber \\
\Lambda (1810;1/2^+) &\simeq &    
\lbrace 0^+\rbrace 
\otimes \lbrack \lbrace {1\over 2},0\rbrace \oplus \lbrace 0,{1\over 2}
\rbrace\rbrack  ,\nonumber \\
\Lambda(1820;5/2^+) &\simeq &    
\lbrace 0^+\rbrace \otimes \lbrack \lbrace {5\over 2},0\rbrace \oplus 
\lbrace 0,{5\over 2}
\rbrace\rbrack  ,\nonumber \\
\Lambda(1830;5/2^-) &\simeq &    \lbrace 0^-\rbrace  
\otimes \lbrack \lbrace {5\over 2},0\rbrace \oplus \lbrace 0,{5\over 2}
\rbrace\rbrack \, . \label{parduhy} 
\end{eqnarray}
In other words, in the $\Lambda $ hyperon sector the parity duplication
occurs for the first time at around 1800 MeV. A possible interpretation of 
this phenomenon is that at that scale the strange quark has already accumulated 
enough energy to move sufficiently far away from the nonstrange quarks, giving 
rise to a reflection--asymmetric nucleon shape. That such a shape is responsible 
for the occurrence of parity doublets in baryon spectra was to our knowledge
initiated by D\"onau and Reinhardt.\cite{DoRe} Later on, extensive work
in that direction was performed, e.g., by Iachello as well as by Bijker and 
Leviatan.\cite{Iachello} For the reasons given above, the $\Lambda $ hyperon 
spectrum above 1800 MeV presents itself as patterned after the classification 
schemes of Refs.~\refcite{DoRe} and \refcite{Iachello} rather than after the 
O(1,3)$\otimes $SU(2)$_I$ scheme. Therefore, for $\Lambda $ hyperons the 
restoration of chiral symmetry takes place above 1800 MeV. 
Similar patterns have been reported by Balachandran and Vaidya\cite{Vaidya}
in the spectra of heavy flavor ($cbu$) and ($cbd$)  baryons.
There, it was shown that once heavy flavor quarks participate in the 
baryon structure, the baryon shape becomes reflection--asymmetric,
thus giving rise to parity doublet patterns.

\vspace*{1pt}\textlineskip	
\section{Propagating Two--Resonance Clust\-ers
and Iso\-lat\-ed\- Spin--3/2 States}
\vspace*{-0.5pt}
\noindent
To find the relativistic equation of motion for a given
$\lbrace j,j \rbrace $
representation of the Lorentz group, one has to construct the
corresponding projection operator. For that one has to find the 
$(2j+1)(2j+1)$ matrices $D^{jj}(b(\vec{p}\, )) $ that carry the boost 
$b(\vec{p}\, ) $ from the rest frame to the system where the particle 
is moving with momentum $\vec{ p}\, $ as well as the respective matrix 
of the three--space inversion ${\cal P}$. The projector onto
$\lbrace j,j\rbrace $ is then expressed as:\cite{Kirchbach}
\begin{equation}
\Pi (p)^{j,j} =
D^{j,j}(b(\vec{p}\, )) {1\over 2}
(\eta_p{\cal P}  +1) \left( D^{j,j}(b(\vec{p}\, )\right) ^{-1}\, .
\label{boost}
\end{equation}
Here, $\eta_p$ is the parity of the vacuum where
the $\lbrace j,j \rbrace $ multiplet is built upon.
With that, the projector $\left( \Pi^{RS}_{\mu \nu }(p)\right)$ onto 
the 16--dimensional RS--field 
$\lbrace 1/2, 1/2\rbrace \otimes \lbrack \lbrace 1/2,0\rbrace 
\oplus \lbrace 0,1/2\rbrace \rbrack $
(subsequently denoted by $\Psi_\mu (p)$ ) is found as
\begin{equation}
\Pi^{RS}_{\mu\nu }(p) =
{{p^\lambda\gamma_\lambda +M} \over {2M}} 
\left(g_{\mu\nu} -{{p_\mu p_\nu }\over M^2}\right)\, ,
\label{RSProj}
\end{equation}
with $\gamma_\lambda $ being the standard Dirac matrices.
The field $\Psi_\mu (p)$ represents a Lorentz vector with
Dirac spinor components. From that, the corresponding relativistic 
equation of motion is obtained as
\begin{equation}
\Pi^{RS}_{\mu\nu} (p) \Psi^\nu (p)  = \Psi_\mu (p) \, .
\label{RS}
\end{equation}
Note that a polar vector $\Psi (p)_\mu$ collects the 
S$_{2I, 1}$, P$_{2I, 1}$, and D$_{2I,3}$ resonances, while
an axial one unites the P$_{2I, 1}$, S$_{2I, 1}$, and P$_{2I, 3}$
states, respectively.
This shows that the redundant components of the RS field are not,
as often said, unphysical ones.
Equation (\ref{RS}) is automatically satisfied if the Dirac equation
is valid for each component of the field $\Psi_\mu $, on the one side,
\begin{equation}
{1\over {2M}} \left( p^\lambda\gamma_\lambda  +M\right)\Psi_\mu (p) =  
\Psi_\mu (p) \, , \label{RS_Dirac}
\end{equation}
and if $\Psi_\mu $ satisfies 
\begin{equation}
\left(g^{\mu\nu } - { {p^\mu p^\nu}\over M^2} \right) \Psi_\nu (p) 
=  \Psi^\mu (p) \, ,
\label{RS_Spin1}
\end{equation}
on the other side.
Equation (\ref{RS_Spin1}) is equivalent to 
Proca's equation $p^\mu\Psi_\mu (p) = 0 $ and is usually considered 
as a supplementary condition to Eq.~(\ref{RS_Dirac}). 
It eliminates one of the four degrees of freedom of the 
Lorentz vector $\Psi_\mu $ (which can be considered to be the time 
component) and ensures mapping of the 
selfconjugate Lorentz multiplet $\lbrace 1/2,1/2\rbrace $  onto the
bi--vector $\lbrace 1,0 \rbrace \oplus \lbrace 0,1\rbrace $.
{}For concreteness, we here consider the case of a polar Lorentz 
vector. In that case it is the P$_{2I,1}$ resonance that is removed 
by means of Proca's equation.  
The remaining 12 components unite the S$_{2I,1}$ and D$_{2I, 3}$ states.
Therefore, the covariant object,
\begin{equation}
{\cal S} _{\mu\nu} (p) = {\Pi^{RS}_{\mu\nu} (p) \over {p^2-M^2}}
= { { (p^\lambda\gamma_\lambda +M) \left(g_{\mu\nu} -
{{p_\mu p_\nu }\over M^2}\right)}\over
{2M(p^2-M^2) }}\, ,
\label{RSProp}
\end{equation}
describes the propagator of the cluster state containing
the resonances D$_{2I,3}$ and S$_{2I,1}$ considered as mass degenerate.
This is quite a reasonable approximation as the splitting between these
states is only 15 MeV for the isodoublet, and only 50 MeV for the isoquadruplet. 
The P$_{2I,1}$ resonance can be considered separately by means of the Dirac 
equation. This way, the problem of the mass splitting within the 
$\lbrace 1/2,1/2\rbrace $ multiplets is solved satisfactorily.

The scheme outlined  above can be extended to multi--resonance cluster 
states. However, the description of isolated higher--spin
$\lbrace J,0\rbrace\oplus\lbrace 0,J \rbrace$ states 
like the two lowest $\Delta $ resonances at 1232 MeV and 1600 MeV, 
respectively, or the parity duplicated higher--spin states observed
in the $\Lambda $ hyperon spectrum, still remains problematic.
Recall that the final isolation of the spin--3/2 component from 
Eqs.~(\ref{RS_Dirac}) and (\ref{RS_Spin1}) is performed by means of a 
second subsidiary condition on $\Psi_\mu (p)$, taken as\cite{Lurie}
\begin{equation}
\gamma^\mu \Psi_\mu (p) =  0\, .
\label{2nd_sc}
\end{equation}
In doing this, the difficulty arises that the subsidiary conditions, 
in particular Eq.~(\ref{RS_Spin1}), are fulfilled only on--mass shell, 
so that off--mass shell the separation between the spin--1/2 and the 
spin--3/2 fields is no longer guaranteed.\cite{Moldauer}
This shows up in the appearance of arbitrary off--shell parameters both 
in the $\Delta $ propagator and the $N\Delta\gamma $ vertex.\cite{Bennm} 

An alternative linear relativistic equation of motion for spin--3/2 fields 
can be obtained in coupling a Dirac spinor  directly to
a 6--dimensional bi--vector rather than to a 4--vector:
\begin{equation}
\Psi_a \simeq
\lbrack \lbrace 1,0\rbrace \oplus \lbrace 0,1\rbrace \rbrack
\otimes 
\lbrack \lbrace {1\over 2},0\rbrace \oplus \lbrace 0, {1\over 2}\, 
\rbrace \rbrack \, , \qquad  a = 1,...,6\, .
\label{bivec}
\end{equation}
In the following, Latin indices will be used to label the 
bi--vector components. 
This 24-dimensional representation, which has been
mapped onto a bi--vector with Dirac spinor components, 
describes a doubled two--resonance spin--3/2 and spin--1/2 cluster 
together with the corresponding anti--cluster. 
The doubling accounts for covering of space--time reflection operations.
Equation (\ref{bivec}) was originally proposed by Fushchich and 
Nikitin (FN).\cite{FuNi} Here we will separate the Dirac from the bi--vector 
indices rather than operate with the complete 24 dimensional FN field as was done 
in the original paper.\cite{FuNi} The advantage of our scheme will become clear in 
constructing interaction Lagrangians.
Now, each component of $\Psi_a $ satisfies the Dirac equation,
\begin{equation}
(/\!\!\!p -M)\Psi_a (p)  = 0\, , \qquad a=1,...,6 \, ,
\label{bivec_bisp}
\end{equation}
with $/\!\!\!p:=p^\mu\gamma_\mu $.
In order to obtain a description of the spin--3/2 field alone,  
a covariant subsidiary condition 
has to be imposed on the wave function $\Psi_a (p) $ 
that eliminates the spin--1/2 component.
This subsidiary condition can be constructed 
directly from the Pauli--Lubanski vector,\cite{ItzZu}
\begin{equation}
(W_\mu)_{ab } = -{1\over 2} 
\epsilon_{\mu\nu\rho\tau} (S^{\nu\rho })_{ab } p^\tau\, ,
\qquad a,b=1,...,6\, ,
\label{PauLu}
\end{equation}
where $(S_{\nu\rho })_{ab }$ are the spin--matrices 
for the composite bi--vector--Dirac spinor representation 
\begin{equation}
(S_{\nu\rho } )_{ab} = 
{1\over 2}\sigma_{\nu\rho } (1_6)_{ab} + 
(\sigma_{\nu\rho }^{(1)})_{ab }1_4\, .
\label{spinrpr}
\end{equation}
Here,  $\sigma_{\mu\nu}=i\lbrack \gamma_\mu ,\gamma_\nu \rbrack /2$
and $\sigma_{\mu\nu}^{(1)} =i\lbrack \gamma_\mu^{(1)} ,
\gamma_\nu^{(1)} \rbrack /4 $, respectively,
where the $6\times 6$ unit matrix acts onto the bi--vector, while the 
$4\times 4$ unit matrix acts within the Dirac spinor space. The matrices
$\gamma^{(1)}_\mu $ are introduced as 
\begin{equation}
\gamma_0^{(1)} = \left(
\begin{array}{lr}
0_3&-1_3\\
-1_3&0_3
\end{array}\right)\, , 
\label{gamma0}
\end{equation}
where $1_3$ and $0_3$ denote in turn the three-dimensional unit and 
zero matrices, while the remaining 6--dimensional spin--1 
$\vec{\gamma} \, ^{(1)}$ matrices  are defined as 
\begin{equation}
\vec{\gamma }\, ^{(1)} =2
\left(
\begin{array}{lr}
0_3 & \vec{S\, }\\
-\vec{S\, } &0_3 
\end{array}\right)\, ,
\label{gamma1_vec}
\end{equation}
with 
\begin{equation}
(S_\alpha )_{\beta \eta } = -i \epsilon_{\alpha \beta \eta }\, ,
\qquad \alpha ,\beta , \eta =  1,2,3\, .
\label{spin1matr}
\end{equation}
At rest the eigenvectors to $S_3$ are given by
\begin{equation}
\vec{\epsilon}_1 = {1\over \sqrt{2}} \left(
\begin{array}{c}
1\\
i\\
0
\end{array}\right)\, , \qquad
\vec{\epsilon}_2 = {1\over \sqrt{2}} \left(
\begin{array}{c}
1\\
-i\\
0
\end{array}\right)\, , \qquad
\vec{\epsilon}_3 =  \left(
\begin{array}{c}
0\\
0\\
1
\end{array}\right)\, .
\label{basis1}
\end{equation}
Note that the $\gamma^{(1)}_\mu $ matrices do not
constitute a Clifford algebra, but they possess well defined 
anti--commutators as the three matrices $S_\alpha $ from 
Eq.~(\ref{spin1matr}) satisfy the su(2) algebra. Finally, we wish to 
introduce the notation for the bi--vector--Dirac spinor as
\begin{equation}
\Psi = \left(
\begin{array}{l}
\vec{\phi }\\
i\vec{\chi }
\end{array}\right)\, ,
\label{MaxDir}
\end{equation}
where $\vec \phi $ and $\vec \chi $ are three-dimensional vectors
with Dirac spinor components. 
At rest and within a basis where the index $a$ counts the angular
momentum states $JM$ one finds
\begin{eqnarray}
\phi_{JM} &=& \sum_{\stackrel{m=0,\pm 1}{s_3=\pm 1/2}}
(1m_1{1\over 2}s_3|JM)\epsilon_{1m}u_s\, ,\\
u_{{1\over 2}} =
\left(
\begin{array}{c}
1\\
0
\end{array}\right) \, , &&
u_{-{1\over 2}} =
\left(
\begin{array}{c}
0\\
1
\end{array}\right) \, . 
\label{rest_spinors}
\end{eqnarray}
Now, the squared Pauli--Lubanski vector is calculated as
\begin{equation}
(W_\mu W^\mu )_{ab}= 
{1\over 2} (S_{\mu\nu })_{ab' }(S^{\mu\nu })_{b' b }
p_\lambda p^\lambda \, 
-(S_{\mu\nu})_{ab'} (S^{\sigma\nu })_{b'b}p^\mu p_\sigma \, .
\label{w2}
\end{equation}
With that the subsidiary condition to Eq.~(\ref{bivec_bisp}),
which eliminates the redundant spin--1/2 component can be deduced as
\begin{equation}
(W^\mu W_\mu)_{ab } \Psi^s _b(p)  = -p^2 s(s+1)\Psi^s _a (p)  \, ,
\qquad s = {3\over 2}\, .
\label{proj1}
\end{equation}
For $p^2\not= 0$ Eq.~(\ref{proj1}) is equivalently cast into the form 
\begin{eqnarray}
P^s _{ab } (p) \Psi^s _b (p) &= &\Psi^s _a (p) \, ,\nonumber\\
P^s _{ab } (p)  &= &-{1\over {2s}}\left[ 
{1\over p^2} (W^\mu W_\mu )_{ab }+s(s-1)\delta_{ab}\right] \, ,
\quad s={3\over 2}\, ,
\label{FNProj}
\end{eqnarray}
where the projection operator $P^s_{ab } (p) $ onto the subspace
with the {\em maximal } spin $s$ has been introduced.
It turns out to be quite favorable to consider Eq.~(\ref{FNProj})
as the equation of motion while Eq.~(\ref{bivec_bisp}) is considered
as the subsidiary condition. The latter is simply incorporated into 
Eq.~(\ref{FNProj}) in rewriting $P^s (p) $ to the equivalent form\cite{FuNi}
\begin{eqnarray}
 P^s_{ab} (p) \Psi^s_b (p) &=& \left({ {1+\gamma_5}\over 2} + 
{ {1-\gamma_5}\over 2}
{ {/\!\!\!p} \over M }\right) 
P^s_{ab} (p) \Psi^s _b (p) \nonumber\\
&=& 
{ {/\!\!\!p + M}\over {2M} }(1+\gamma_5)(-1){1\over {4s}}
    {\Big[}(S^{\mu\nu}S_{\mu\nu })_{ab} +2s(s-1)\delta_{ab}\nonumber\\
& &-
   2(S_{\mu\nu}S^{\sigma\nu })_{ab} {{p^\mu p_\sigma}\over p^2}{\Big]}
   { {/\!\!\!p} \over M } \, \Psi^s_b (p) \, , \nonumber\\
P^s (p) ^2 = P^s (p) \, , \quad P^{s-1} (p)  &= & 
1- P^s (p) \, ,\quad P^s (p)\, P^{s-1} (p) =0\, ,\,\,\, s={3\over 2}\, ,
\label{maineq}\end{eqnarray} 
where $\gamma_5 = i\gamma_0\gamma_1\gamma_2\gamma_3$, and use of 
Eqs.~(\ref{w2}) and (\ref{proj1}) was made. From Eq.~(\ref{maineq}) the 
propagator of the spin--3/2 field is deduced as the following matrix:
\begin{eqnarray}
{\cal S}^{{3\over 2}}(p) &= & { {P^{{3\over 2}} (p) }\over {p^2-M^2}}\, , 
\nonumber\\ S^{{3\over 2}} (p)
&=&-{ {(/\!\!\!p+M)(1+\gamma_5){\Big[} (S^{\mu\nu}S_{\mu\nu}+ {3\over 2}
-{2\over p^2} S_{\mu\nu}S^{\sigma\nu} p^\mu p_\sigma {\Big]}
/\!\!\!p }\over {12M^2 (p^2-M^2)} }\, . 
\label{FNProp}
\end{eqnarray} 

It is easily verified that Eq.~(\ref{maineq}) is obtained from the 
following Lagrangian
\begin{eqnarray}
{\cal L} &= & \bar{\Psi}^s_a \, (/\!\!\!p +M)(1+\gamma_5) 
     {\Big[} (S^{\mu\nu} S_{\mu\nu })_{ac} +2s(s-1)\delta_{ac} \nonumber\\
& & -2(S_{\mu\nu} S^{\sigma\nu })_{ac} { {p^\mu p_\sigma}\over M^2} 
{\Big]} { {/\!\!\!p}\over M }\Psi^s_c +8M\, s\ \bar{\Psi}^s_d\Psi^s_d\, ,
     \quad s={3\over 2}\, , \label{lagr}
\end{eqnarray} 
where $p^2 =M^2$ has been set. The introduction of a minimal 
coupling of the field $\Psi $ to the photon via 
$p_\mu \to p_\mu - eA_\mu $ preserves the form (and therefore the 
compatibility) of both the Eqs.~(\ref{bivec_bisp}) and (\ref{maineq}). 
As long as Eq.~(\ref{FNProj}) is not restricted to on--mass shell only,
the separation of the spin--3/2 and spin--1/2 components does not
cause any more problems.
The interacting propagator is now obtained from 
Eq.~(\ref{FNProp}) by replacing $/\!\!\!p$ through
$/\!\!\!\pi :=(p^\mu-eA^\mu)\gamma_\mu $. In\cite{FuNi} the frame 
independence of the eigenvalue of the Pauli--Lubanski vector 
was exploited to replace the second term of Eq.~(\ref{w2}) by its 
equivalent in the rest frame, $-S^{\mu\nu}S_{\mu \nu }/4$. For that case, 
the proof was given, that the spin--3/2 particle does not any longer move
with superluminal velocity in an external magnetic field.

Now, in mapping the bi--vector onto a totally antisymmetric field tensor 
$V_{\lbrack \mu\nu \rbrack } $ according to
\begin{eqnarray}
V_{0 \alpha } &=&
\phi_\alpha \, , \qquad \alpha =1,2,3\, , \nonumber\\
V_{12  }  = \chi_3\, ,\qquad V_{23} &= &\chi_1\, ,
\quad   V_{31} = \chi_2\, ,
\label{Vtensor}
\end{eqnarray}
one can map the 24--dimensional representation $\Psi_a (p)$
onto a 2nd rank antisymmetric Lorentz tensor with Dirac spinor
components according to
\begin{equation}
\Psi_a (p) \simeq \psi_{\lbrack \mu \nu\rbrack }(p) \, ,
\label{antisym}
\end{equation}
where the brackets `$\lbrack\, \, \,  \rbrack $' have been used to 
emphasize the antisymmetrization. The favor of our notation over that
of the original paper\cite{FuNi} is the that the spin--3/2 field is 
described as a fully covariant object. Now the $ND_{2I, 3} \gamma $ 
vertex in coordinate space can be constructed as
\begin{equation}
{\cal L}_{ND_{2I,3}\gamma } = g \bar {\psi}_{ \mu\nu }\psi_N F^{\mu\nu }
+ g_1\bar\psi^{\mu\nu}
\sigma_{\mu\lambda}\sigma_{\nu\eta}\psi_N F^{\lambda\eta}\, , 
\label{NDPhot}
\end{equation}
where $\psi_N$ stands for the Dirac spinor of the nucleon,
$F^{\mu\nu}$ is the electromagnetic field strength tensor,
while $g$ and $g_1$ are the coupling strengths. 
Other couplings as
\begin{equation}
\bar{\psi}^{\mu\nu}\sigma_{\mu\nu}\sigma_{\rho\tau}\psi_N
F^{\rho\tau }\, ,
\end{equation}
will not contribute as $\sigma_{\mu\nu}\psi^{\mu\nu }$
vanishes. Indeed, at rest, this condition reduces to
$\vec{\sigma }\cdot \vec {\phi } =0$ and ensures validity of
Eq.~({\ref{FNProj}) in the rest frame, where it takes the form
\begin{equation}
\left(\vec{S} +{1\over 2}\vec{\sigma }\right)^2\vec{\phi} 
= {15\over 4}\vec{\phi }\, .
\label{w2rest}
\end{equation}
As long as Eq.~(\ref{FNProj}) is valid off--mass shell too,
the vertex constructed in Eq.~(\ref{NDPhot}) does not contain any 
off--shell parameters. 
The $ND_{3I, 3}\pi $ vertex in coordinate space can be constructed in 
analogy to Eq.~(\ref{NDPhot}) by replacing $F_{\mu\nu }$ by 
$U_{\mu\nu}=\lbrack U^\dagger \partial_\mu U, U^\dagger \partial_\nu U
\rbrack $ with $U$ being the well known dimensionless quaternion chiral 
meson field $U=(\sigma + i\vec{\pi }\cdot \vec{\tau })/f_\pi$ underlying 
the Skyrme model.

\vspace*{1pt}\textlineskip	
\section{Summary}
\vspace*{-0.5pt}
\noindent
In the present paper it was shown that the clustering
in baryon spectra pointed out by H\"ohler\cite{Hoehler}
is well interpreted in terms of multiplets
of the Lorentz--isospin group O(1,3)$\otimes $SU(2)$_I$.
{}For nonstrange baryons these clusters are of the type 
$\lbrace j,j\rbrace \otimes \lbrack \lbrace 1/2,0\rbrace \oplus
\lbrace 0, 1/2\rbrace \rbrack \otimes \lbrace I\rbrace $ with 
$n=2j+1$ even. For the $\Lambda $ hyperons with masses below 
1800 MeV the Lorentz clusters have been shown to be of the type 
$\lbrace 1,0\rbrace \oplus \lbrace 0,1\rbrace\otimes 
\lbrack\lbrace 1/2,0\rbrace \oplus \lbrace 0,1/2\rbrace \rbrack $
while above 1800 MeV they are parity-duplicated $\lbrace J,0\rbrace 
\oplus \lbrace 0, J \rbrace $ Weinberg--Ahluwalia states. 
The suggestion was made that in evaluating Feynman graphs in such 
processes as meson production off protons, the propagation of 
resonance clusters as intermediate states might be more appropriate and 
easier to handle as compared to the propagation of isolated 
higher--spin states. Clusters corresponding to Lorentz multiplets with 
$n=2j+1 $ odd have not been observed so far. At the present stage it is 
still not clear if their existence is forbidden for some dynamical reasons 
or if they have to be viewed as missing clusters because of the
property to decouple from the $\pi N$ decay channel. The possibility to 
observe missing resonances in other than the $\pi N$ decay channel
was discussed in a recent work by Capstick and Roberts.\cite{Capstick} 

In addition, for nonstrange baryons the Nambu--Goldstone mode of chiral 
symmetry was shown to extend only to the first excited P$_{2I,1}$, 
S$_{2I,1}$, and D$_{2I,3}$ states (with $I= 1/2$, and $I=3/2$). The next 
higher Lorentz multiplets with $n= 2j+1 > 2 $ were shown to be built upon 
a $0^-$ vacuum and hence, chiral symmetry is restored there.
{}For $\Lambda $ hyperons chiral symmetry is restored above 1800 MeV. 

As a promising cand\-id\-ate for des\-crib\-ing  the quark 
dy\-namics underlying the O(1,3)$\otimes $SU(2)$_I$ spectrum--generating 
algebra of Ref.~\refcite{Kirchbach} one may consider the covariant 
quark--diquark model based on solving the Bethe--Salpeter equation in the 
rapidly converging O(4) basis of the Gegenbauer polynomials as outlined 
in.\cite{Kus}

{}Finally, it was demonstrated that the description of spin--3/2 
particles in terms of a 2nd rank antisymmetric Lorentz tensor with Dirac 
spinor components does not contain any off--shell parameters and 
avoids the main difficulties of the description based 
upon a 4--vector with Dirac spinor components.

\vspace*{1pt}\textlineskip	
\section{Acknowledgements}
\vspace*{-0.5pt}
\noindent
Valuable correspondence with Anatoly Nikitin is gratefully acknowledged.  
This work was supported by the Deutsche Forschungsgemeinschaft (SFB 201).

\nonumsection{References}

\end{document}